\documentclass[aps,prb,twocolumn,groupedaddress]{revtex4-1}
\usepackage{graphicx, subfigure}

\begin{document}


\title{Intrinsic Defects and Dopability of Zinc Phosphide}


\author{Steven Demers}
\email{steved@caltech.edu}
 \affiliation{Department of Applied Physics and Materials Science, California Institute of Technology, Pasadena, California 91125, USA}
\author{Axel van de Walle}
\affiliation{Department of Applied Physics and Materials Science, California Institute of Technology, Pasadena, California 91125, USA}

\date{\today}

\begin{abstract}
Zinc Phosphide ($Zn_3P_2$) could be the basis for cheap and highly efficient solar cells. Its use in this regard is limited by the difficulty in n-type doping the material.  In an effort to understand the mechanism behind this, the energetics and electronic structure of intrinsic point defects in zinc phosphide are studied using generalized Kohn-Sham theory and utilizing the Heyd, Scuseria, and Ernzerhof (HSE) hybrid functional for exchange and correlation.  Novel 'perturbation extrapolation' is utilized to extend the use of the computationally expensive HSE functional to this large-scale defect system.  According to calculations, the formation energy of charged phosphorus interstitial defects are very low in n-type $Zn_3P_2$ and act as 'electron sinks', nullifying the desired doping and lowering the fermi-level back towards the p-type regime.  This is consistent with experimental observations of both the tendency of conductivity to rise with phosphorus partial pressure, and with current partial successes in n-type doping in very zinc-rich growth conditions.
\end{abstract}


\maketitle

\section{\label{intro}Introduction}
Zinc Phosphide has great potential as a photovoltaic material; it absorbs strongly in the visible spectrum ($>10^4-10^5 cm^{-1}$), has long minority carrier diffusion lengths (5-10 $\mu m$), and a direct, almost ideal bandgap ($\sim$1.5eV)\cite{kimball}.  Both zinc and phosphorus are Earth-abundant elements, greatly aiding their widespread use.  However, no practical means of creating $Zn_3P_2$ crystals with n-type doping has been found.  This has prevented the typical p-n homojunction solar cells and current implementations instead rely upon metal-semiconductor junctions or p-n semiconductor heterojunctions\cite{Faa}.  The best results are currently with p-$Zn_3P_2$/Mg Schottky diode where the maximal open-circuit voltage ($\sim$0.5eV) is the limiting factor on the efficiency of these devices (currently 6\%)\cite{Bhushan}.

Differences in growth environment has been shown to affect the electrical properties significantly\cite{zdan}.  Specifically, resistivity ranging over 100 to $10^5 \Omega$-cm have been measured in single and polycrystalline samples.  Such a large range of values may point to an intrinsic defect mechanism dominating carrier concentrations.  Despite many years of research, the exact nature of the various intrinsic defects as well as their small-scale structure and properties are only partially known.  So, as a step towards understanding the n-type doping difficulties, it is natural to first explore the role of intrinsic defects in this system.  

Density Functional Theory (DFT) is the method of choice for studying the electronic structure of defects in semiconductors.  It is the only \emph{ab initio} approach currently tractable for calculation of the energetics of cells of sufficient size to explore isolated defects (at least on the order of 100's of atoms).  For computational simplicity, the most common exchange-correlation functionals utilized are the local-density approximation (LDA) and generalized-gradient approximation (GGA).  However, while the defect-cell total energies are typically well-described by LDA or GGA, the well-known bandgap problem of these methods poses problems in the determining the precise electronic structure of the defects themselves.  Subsequently, this can lead to different predictions of defect stability amongst calculations even on identical systems.  For example, recent work of various groups on ZnO has lead to predictions that the oxygen vacancy acts as both a shallow and deep defect\cite{LanyZnO,JanottiZnO}.  Recently, more accurate techniques such as hybrid functionals\citep{HSE} and GW excited-state calculations\citep{GW} yield greatly improved bandgap prediction.  However, these techniques are still too expensive for the large scales required of defect systems.

Regardless of the exact \emph{ab initio} procedure chosen, there is a fundamental problem in studying defects with a necessarily small amount of atoms (even the most efficient approach of LDA-based DFT is limited to on the order of hundreds of atoms); that is, we are imposing a degenerate doping condition, which is seldom the regime of interest (typical defect or dopant concentrations rarely exceed parts-per-million).  As such, care must be taken to correct for interactions that are artifacts of this incorrectly high defect concentration such as image charges and spurious hybridization.  In the following section we detail the methods that allow us to compute defect levels and energetics separate from these effects.

$Zn_3P_2$ exists in two phases, tetragonal and cubic.  We have investigated the intrinsic point defects of the tetragonal phase, which is of primary concern because it is the room-temperature phase.  These defects include zinc vacancy ($V_{Zn}$), phosphorus vacancy ($V_P$), zinc interstitial ($Zn_i$), and phosporus interstitial ($P_i$).  The antisite defects were also studied, but their formation energies are such that they would be unstable to dissociating into a vacancy-interstitial pair, thus they are excluded from the following discussion.  What we find is that the formation energy of the charged acceptor defects, especially $P_i$, is small enough at even moderate n-type conditions to form enough compensating defects as to completely neutralize the desired extrinsic doping.
\newline
\newline
\newline

\section{\label{comput}Methodology}
Of central concern is the stability of the intrinsic defects in $Zn_3P_2$ and their effect on the fermi level.  The stability of a defect is largely determined by its formation energy; in the supercell formalism put forth by Lany and Zunger\cite{mondo} the formation energy of a defect with charge q is given by a sum of three terms:\begin{equation}
\Delta H_f=[E_{D}-E_H]+q(E_V+\Delta E_f)+\sum_\alpha n_\alpha (\mu_\alpha^o+\Delta \mu_\alpha) \label{FE}\end{equation}
$E_H$ and $E_D$ are the total energy of the perfect host and host+defect supercells respectively.  The first term on the right is the difference in bond energy brought about by the defect.

The second term in (\ref{FE}) represents the energy cost of exchanging electrons with the 'electron reservoir'.  $E_V$ is the reference energy of the reservoir and is the price we pay for removing an electron from the top of the valence band (ie. the energy of a hole at the Valence Band Maximum)\citep{persson}.  Consequently, equation (\ref{FE}) describes the energetics of forming a defect while conserving charge \citep{mondo}.  The calculated total energy ($E_H$) of the system follows Janak's Theorem\citep{Lanymod}:
\begin{equation}
\frac{dE_H(n_i)}{dn_i}=e_i
\label{janak}\end{equation}
where $n_i$ is the occupation of the highest occupied state $i$ with eigenvalue $e_i$.  For an infinite system $e_i$ is identical to $E_V$.  Thus, $E_V$ can be calculated as the energy difference between a host and a host+hole cell in the limit that the number of electrons (N) tends to infinity:
\begin{equation}
\lim_{N\to\infty} [E_H(N) - E_H(N-1)] = E_V
\label{hole}\end{equation}
As a practical matter, a good approximation can be attained for relatively small systems - in the present work the difference between a neutral perfect supercell and a supercell and a hole ($E_H(0)$ - $E_H(+)$) is used.  We now have a good approximation for the electron chemical potential at the VBM.  Finally, $\Delta E_f$ is the additional energy of electrons in our system above the VBM and is the proxy for specifying the doping regime (p or n-type) of the bulk.  

The crystal growth environment affects the formation energy via the chemical potentials ($\mu_\alpha=\mu_\alpha^o+\Delta \mu_\alpha$) in equation (\ref{FE}).  These represent the energy cost of exchanging atoms with the chemical reservoir.  By convention, the formation energies are defined in relation to the standard states of the constituents of a system.  Thus, the elemental chemical potentials are broken into two parts:  $\mu_ \alpha^o$ is the chemical potential of the standard state of the element and $\Delta \mu_\alpha$ is the chemical potential of the element in relation to the standard state.  The growth conditions are reflected in $\Delta \mu_\alpha$ (ie. a maximally rich growth environment of a certain element would have $\Delta \mu_\alpha$ = 0), it becomes more negative for lower concentrations of an element during crystal formation.  The chemical potentials are added or subtracted from the formation energy according to the number of atoms of a certain species is deposited or withdrawn from the growth reservoir ($n_ \alpha$); it is +1 if an atom is added (ie. vacancy defects), if we remove an atom it is -1 (ie. interstitial defects).

Thus, the formation energy is determined as a function of fermi-level ($\Delta E_f$) for a specific growth condition (dictated through $\Delta \mu_\alpha$).  Assuming equilibrium growth conditions, the chemical potentials are restricted to values that maintain a stable compound and don't permit competing phases to exist.  For $Zn_3P_2$ we have the following constraints:

\begin{figure}[t]
\caption{\label{alchem}$Zn_3P_2$ allowed chemical potentials.  The red and green lines are the stability limits for $Zn_3P_2$ and $ZnP_2$ respectively.  The dark purple region corresponds to the range of chemical potentials where we are assured of forming zinc phosphide crystals.}
\includegraphics[trim=1.5cm 1.2cm 2cm 3cm, clip=true,width=80mm]{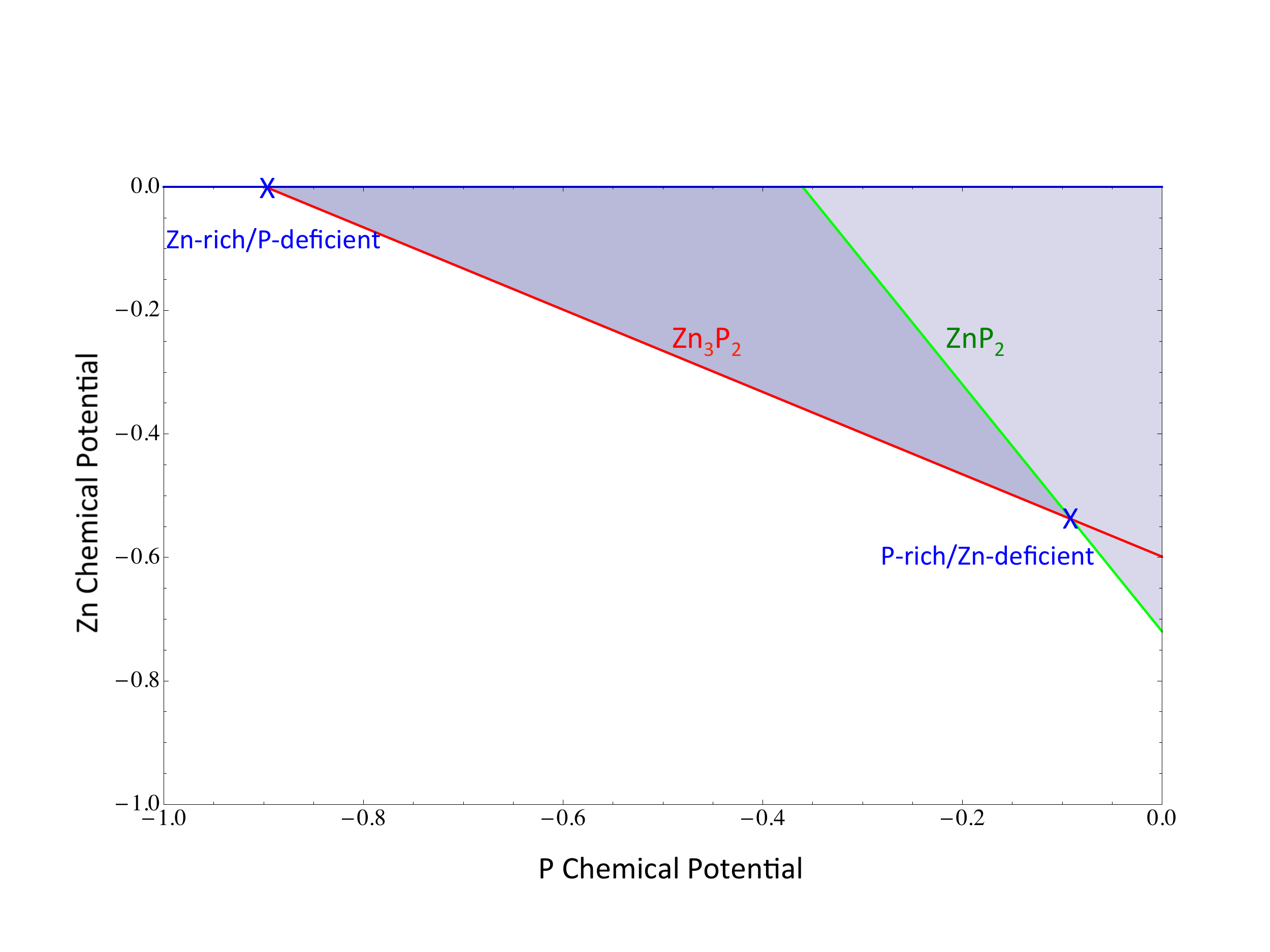}
\end{figure}

In order not to precipitate the elemental form of Zn or P, we must have:
\begin{equation}
\mu_{Zn} \leq \mu_{Zn}^o ; \; \mu_{P} \leq \mu_{P}^o\end{equation}
or, equivalently:
\begin{equation}
\Delta\mu_{Zn} \leq 0 ; \; \Delta\mu_{P} \leq 0\end{equation}
where $\Delta \mu_{Zn}$=0 would define the maximumly-rich Zn growth condition.\newline

The stability condition for $Zn_3P_2$ is:
\begin{equation}
3\Delta\mu_{Zn}+2\Delta\mu_{P} \leq \Delta H_f(Zn_3P_2)\end{equation}
The only other competing phase is $ZnP_2$ (phosphorus-rich phase) and avoiding its formation yields the following restriction:
\begin{equation}
\Delta\mu_{Zn}+2\Delta\mu_{P} \geq \Delta H_f(ZnP_2)\end{equation}
There is a the further issue of incomplete error cancellation in DFT when energy differences are taken between chemically dissimilar systems\citep{Lanychem} (such as between compounds and their elemental constituents).  Here we correct the chemical potentials according to [\citenum{Lanychem}] which results in better agreement between predicted and experimental formation energies (1.04 eV non-corrected, 1.79 eV corrected, 1.5 eV experimental).

Combining all the conditions described by Eq 5-7, we can determine an allowed region of chemical potentials where we are assured of forming only $Zn_3P_2$ - see Figure \ref{alchem}.  Later, when we discuss formation energies for a specific growth environment it will be this figure which defines the chemical potentials to use for the Zn-rich/P-poor versus the Zn-poor/P-rich regimes.
\newline

All these calculations are performed within the Kohn-Sham framework (DFT) and utilize the projector augmented pseudopotentials as implemented in VASP\citep{PAW,PAW-VASP,VASP}.  We use valence configurations of 3d4s and 3s3p for zinc and phosphorus respectively with the Perdew, Burke, and Ernzerhof (PBE) potential \citep{PBE}.  GGA as well as hybrid functionals (HSE) for exchange-correlation were used, the specifics of which will be discussed later.  For supercells of 2x2x2 unit cells (320 atoms) energies were calculated with two k-points (0,0,0) and (1/2,1/2,1/2) and a plane-wave basis with a cutoff of 300eV. These settings result in a numerical error on the order of 0.1 eV in the defect formation energy which is on par with the other sources of error (e.g. uncertainty in the bandgap and finite-size corrections).  All calculations were performed considering spin polarization.

Although the use of a supercell geometry within a DFT framework is a common approach for defect calculations, care must be taken to avoid spurious finite-size effects as well as known deficiencies in DFT's ability to model excited state energies.

\subsection{Finite Size Effects}
With current limitations on computing power, DFT calculations are typically restricted to cells on the order of hundreds of atoms.  Even the addition of a single point defect would thereby result in defect concentrations on the order of tenths of a percent, normally describing degenerate conditions.  Usual semiconductor defect concentrations are much more dilute (on the order of parts-per-million), and often result in far different material properties than within the degenerate regime.  Due to the low dielectric strength (and hence poor screening) of $Zn_3P_2$ these effects are especially troublesome in this study.  However, if we are careful about correcting our results, we can still make accurate predictions about what the dilute environment should look like.  The three main finite-size corrections are discussed below.

\subsubsection{\label{imchg}Image Charge Correction}
The drawback of the use of a standard supercell geometry is that defects are periodically and infinitely repeated spatially.  The defect, instead of being surrounded by a large region of perfect bulk crystal as it would be under non-degenerate conditions, is now surrounded by mirror images of itself.  This will result in somewhat frustrated ionic relaxation, though these elastic energy effects tend to be short range and is rarely a problem for even modestly sized cells (there is very little difference in relaxation energies for even a 2x2x2 supercell versus a unit cell of $Zn_3P_2$).  However, when dealing with charged defects we form 'image charges' leading to spurious electrostatic interactions.  These coulombic interactions between the defect and its mirror charges are long-ranged and significant even for large cells.

Corrections for this 'image charge' effect have been the subject of much research, though the most common approach is based on the work of Makov and Payne\cite{makov}.  They considered the charge density to be the contribution of the periodic charge of the underlying crystal structure and the charge density of the aperiodic defect (which is simply the electron density difference between the host and host+defect cells).  The multipole correction to the formation energy is:
\begin{equation}
E_{IC} = \frac{q^2\alpha_M}{2\epsilon L}+\frac{2\pi qQ_r}{3\epsilon L^3}+O(L^{-5})
\end{equation}
where $\alpha_M$ is the supercell lattice-dependant Madelung constant, L is the length of the supercell, $\epsilon$ is the static dielectric constant, and $Q_r$ is the second radial moment of the aperiodic charge density.  The first two terms are the monopole and quadrupole corrections respectively; the quadrupole correction typically $\sim 30\%$ of the monopole term\citep{Lanymod}.

\subsubsection{Potential Alignment Correction}
\begin{figure}[t]
\caption{\label{potal}Potential alignment correction.  Differences in electrostatic potential between perfect and defect cells far from defect area are used to correct charged cell energies.}
\includegraphics[trim=0cm 1cm 1cm 2cm, clip=true,width=80mm]{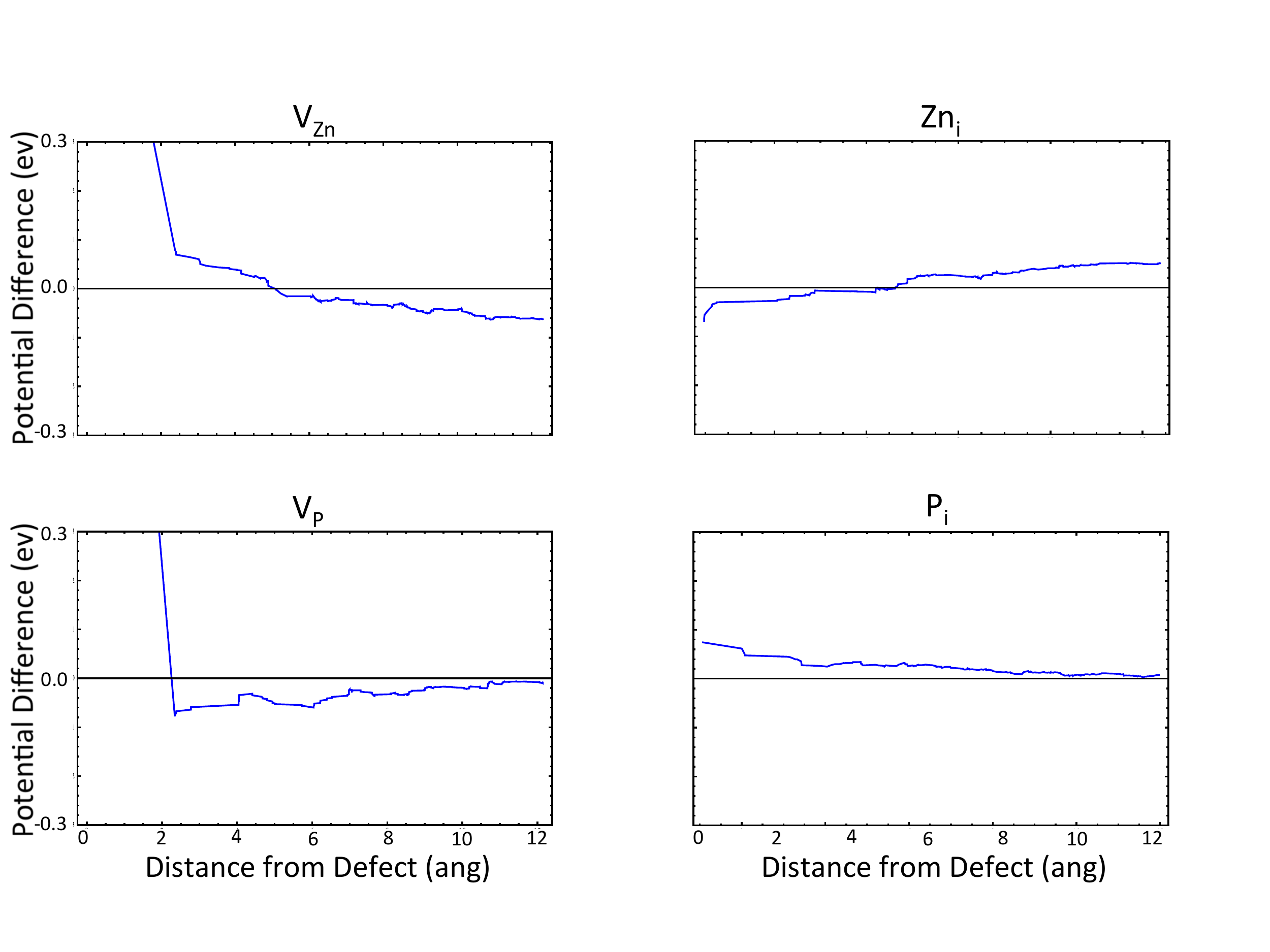}
\end{figure}

In the case of charged defects with periodic boundary conditions there is a violation of charge neutrality, which causes the Coulomb potential to diverge\citep{momentum}.  In momentum-space formalism, one usually sets the G=0 term of the electrostatic and ionic potential ($V_H$(G=0) and $V_I$(G=0)) to zero.  The Kohn-Sham eigenvalues are thus only defined with respect to the average electrostatic potential of the cell.  For neutral systems this arbitrary offset still leads to a well-defined total energy since the electron-electron and ion-ion contributions exactly cancel.  In a charged system, ignoring the G=0 term can be viewed as equivalent to a uniform background charge (jellium) compensating for the net charge - though it is important to note that this only occurs for the potential.  In a charged cell there is now an arbitrary offset to the total energy.  The charged cell energies $E_{D}$ and $E_H(+)$ in Eq. (\ref{FE}) have to be compensated for in order to treat them on an equal footing with the neutral cell\citep{LanyPOTAL}.  The potential-alignment correction is:
\begin{equation}
E_{PA} = q(V_{D,q}^r-V_H^r)
\end{equation}
where the charged defect ($V_{D,q}^r$) and host reference ($V_H^r$) potentials are atomic sphere-averaged electrostatic potentials far from the defect site, q is the charge of the defect- see Figure \ref{potal}.

\subsubsection{Band-filling Correction}
The artificially degenerate doping regime can cause defects to form bands rather than isolated states within the bandgap \citep{moss,burstein}.  For shallow defect states, this incorrect dispersion can cause abnormally large hybridization with the extended band states (either conduction or valence) and subsequently partially populate these bands.  In order to obtain accurate defect cell energies ($E_{D}$) we have to correct for the extra energy in the system due to electron populations at these higher (or lower) energies.  For shallow donors, the correction is: 
\begin{equation}
E_{BF} = -\sum_{n,k} \Theta (e_{n,k} - e_C)(w_k f_{n,k} e_{n,k} - e_C)
\end{equation}
Where $e_{n,k}$ is the k-dependent energy of state at band index n, $e_C$ is the conduction band minimum energy of the defect-free bulk after potential alignment, $f_{n,k}$ is the band occupation, and $w_k$ is the k-point weight. $\Theta$ is the Heaviside step function.

\subsection{Band-gap Error}
The most common exchange-correlation potentials, LDA and GGA, severely underestimates the bandgap of most semiconductors.  This has two damaging effects on defect calculations.  The defect-induced states may lie artificially close in energy to some band states causing excessive hybridization and ambiguity between shallow or deep defect behaviour.  Secondly, the range of electron chemical potentials used to calculate defect formation energies will be too small, possibly incorrectly predicting unstable charged defect states.

Fundamentally, the bandgap error in LDA or GGA is the result of a lack of continuity with respect to the number of electrons in the exchange-correlation potential\citep{discont}.  This in turn leads to self-interaction error (SIE) associated with a bias towards delocalized wavefunctions.  The most common means to correct this situation is to add a Hubbard-like potential to penalize partial state occupancies as in GGA+U.  Unfortunately, there are many equivalent ways to apply this correction (ie. which choice of 'orbitals' to apply this to) yielding the same bandgap.  Since defect levels are sensitive not to the bandgap itself, but to their position relative to the host band states, this ambiguity can result in many different predictions for defect ground states \citep{JanottiZnO,LanyZnO,janotti}.  Furthermore, corrections for SIE are especially important for charged defect calculations as reducing interaction error tends to increase the ionicity of the crystal\cite{ntype} resulting in more ionic relaxation as a defect state is populated and hence greater energy benefit for a charged defect.

Functionals which attempt to correct for SIE have recently emerged.  One of the most robust and computationally tractable is the HSE functional\citep{HSE}.  The exchange-correlation potential is divided into short and long-range components via a screening length parameter.  For the long-range portion, things are unchanged from a typical GGA calculation.  Short-range interactions have a portion of exact Hartree-Fock exchange mixed into the exchange potential, which partially corrects for SIE.  This approach has been shown to greatly increase the accuracy of the bandgap as well as relative band positions for a wide variety of materials\citep{HSEbandgap}.  In this work, the amount of short-range HF-exchange mixing is set to 25\% which is the amount suggested by the adiabatic connection theorem\citep{adiabatic}.  We use a screening length of .1 $\AA^{-1}$ instead of the more typical .2 $\AA^{-1}$ to account for the low dielectric strength, and hence poor screening of $Zn_3P_2$.  This results in close agreement between our calculated bandgap and experiment (see Table \ref{Tab1}).   

However, the added memory requirements associated with the use of the HSE functional for a 2x2x2 supercell of $Zn_3P_2$ (320 atoms) make it computationally intractable with our existing resources (even smaller supercells represent a challenge).  Here we have decided to use the 'perturbation extrapolation' method put forth by Lany, et al. \citep{mondo}  This is based on the idea of expressing the defect-influenced states in the basis of the states of the perfect bulk (assuming these form a complete basis):

\begin{equation}
\Psi_D(r) = \sum_{n,k} A_{n,k} \Psi_{n,k}(r)
\label{def_wave}
\end{equation}    

If we model the bandgap correction of the HSE functional as a perturbation ($H_p$) of the perfect bulk system Hamiltonian ($H_{bulk}$) via a multiplier $\lambda$, the band energies shift:
\begin{equation}
e_{n,k}(\lambda) = \langle \Psi_{n,k} \vert H_{bulk} \vert \Psi_{n,k} \rangle +
\lambda \langle \Psi_{n,k} \vert H_p \vert \Psi_{n,k} \rangle
\end{equation}

\begin{equation}
e_{n,k}(\lambda) = e_{n,k}(0) + \lambda \frac{\partial e_{n,k}(\lambda)}{\partial \lambda}
\end{equation}

If we apply this same perturbation to the defect system (with Hamiltonian $H_{bulk} + H_D$) then we have:

\begin{equation}
e_D(\lambda) = \langle \Psi_D \vert H_{bulk} + H_D \vert \Psi_D \rangle +
\lambda \langle \Psi_D \vert H_p \vert \Psi_D \rangle
\end{equation}

\begin{equation}
e_D(\lambda) = e_D(0) + \lambda \frac{\partial e_D(\lambda)}{\partial \lambda}
\end{equation}

\begin{figure}[t]
\caption{\label{workflow}Perturbation Extrapolation Workflow - assuming that the perfect unit cell GGA forms a complete basis for the supercell defect wavefunctions, we project the GGA defect states onto the GGA perfect states and then use the offsets between the GGA and HSE unit cells to extrapolate HSE supercell behavior.}
\includegraphics[trim=1cm 1cm 6cm 4cm, clip=true,width=60mm]{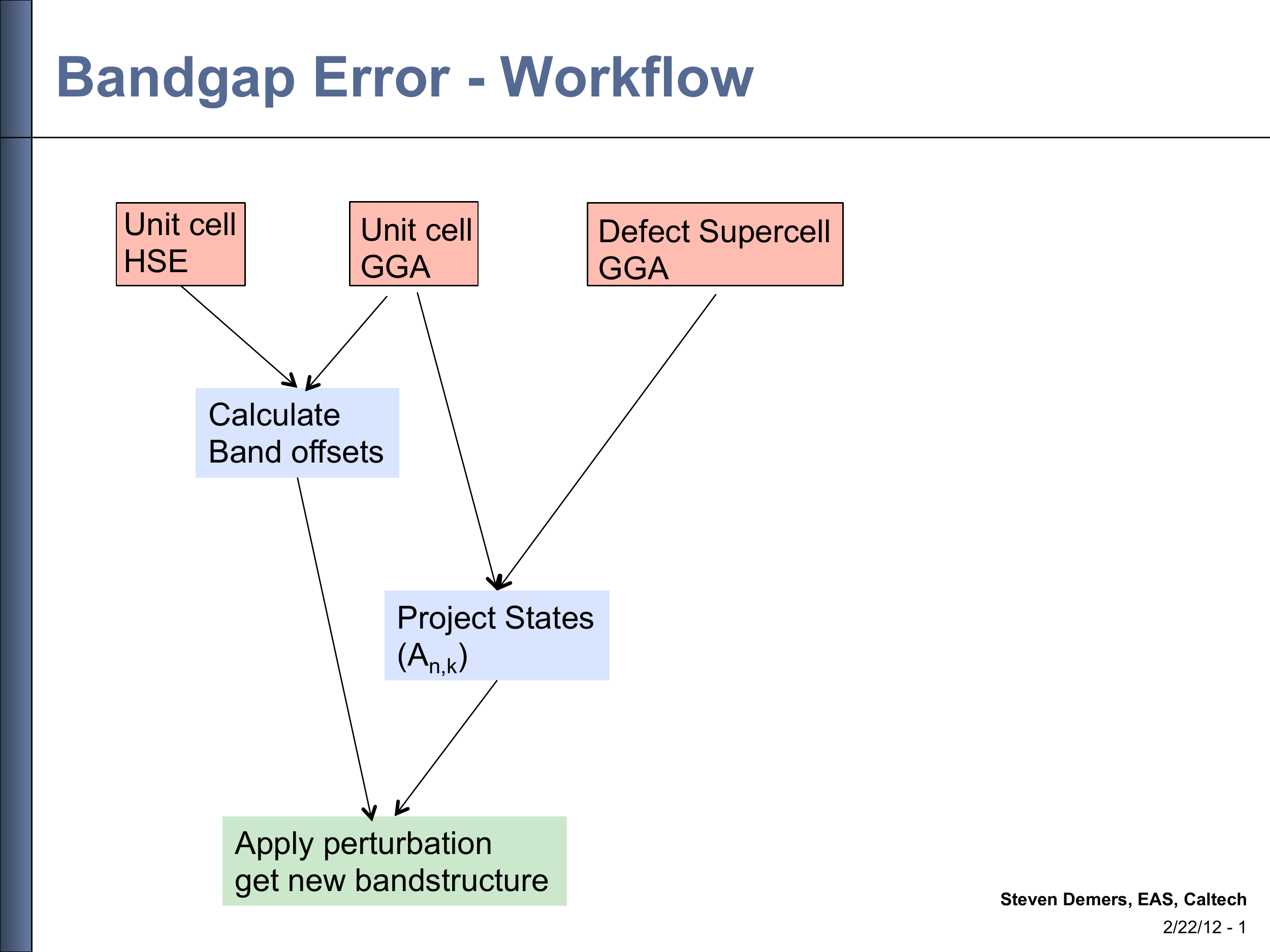}
\end{figure}

Under the assumption of first-order perturbation theory and via Eq. (\ref{def_wave}) we have the final result: 
\begin{equation}
e_D(\lambda) = e_D(0) + \lambda \sum_{n,k} A_{n,k}^2 \frac{\partial e_{n,k}(\lambda)}{\partial \lambda}
\end{equation}
As long as the assumptions of first-order perturbation theory are justified (ie. unchanged wavefunctions upon application of the band-correcting perturbation $H_p$) we would predict the defect states to track the corrections to the perfect bulk states in proportion to the square of the coefficients in the expansion ($A_{n,k}^2$) of Eq. (\ref{def_wave}).  As HSE doesn't significantly affect the dispersion of bands so much as their relative positions to each other, we expect that this approximation should be justified.

There is still a question as to whether the perfect bulk states form a reasonably complete basis to describe the defects.  We are helped here by the fact that the static dielectric strength of $Zn_3P_2$ is low ($\sim 3$) which leads to the tendency of the defects to not be very localized.  The number of bands we need to calculate in the expansion of Eq. (\ref{def_wave}) is thus fairly limited.

The general workflow is shown in Figure \ref{workflow}.  Perfect bulk unit cell wavefunctions are determined with GGA and HSE functionals.  Then a 2x2x2 defect supercell is calculated only with the GGA functional.  The defect supercell GGA wavefunctions are then projected onto the unit cell GGA wavefunctions obtaining $A_{n,k}$.  Each band is then offset by the amounts prescribed by comparing the GGA and HSE unit cells.  We can then make a prediction for what the supercell defect bandstructure would be if we had been able to utilize the HSE functional.

\begin{figure}[t]
\caption{\label{unitcell}Tetragonal $Zn_3P_2$ has a 40-atom unit cell with P42/nmc symmetry.}
\includegraphics[trim=3cm 1cm 3cm 1cm, clip=true,width=50mm]{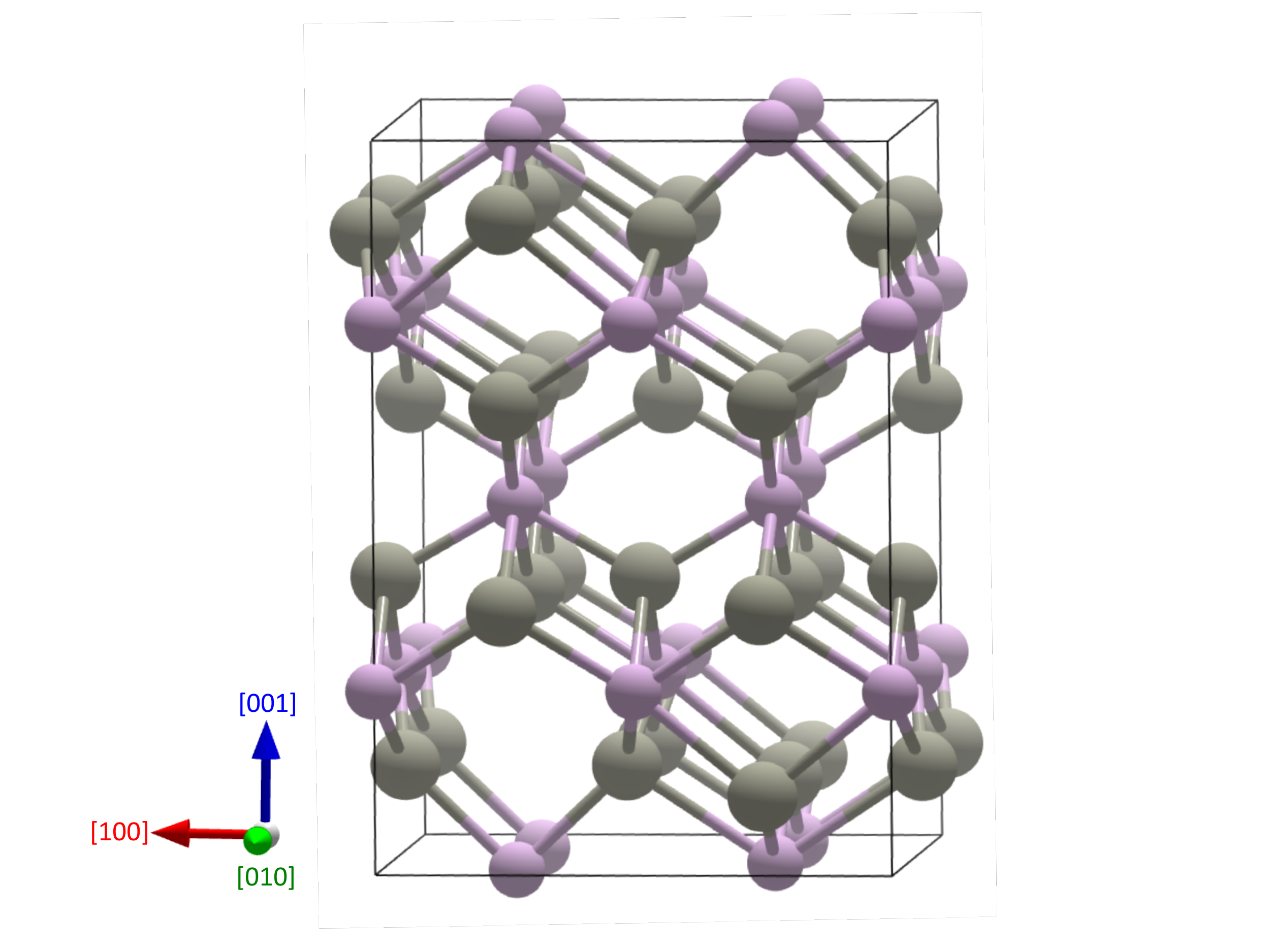}
\end{figure}

\begin{table}[b]
\caption{\label{Tab1}
Calculated lattice constants, bandgap, and heat of formation of tetragonal $Zn_3P_2$ using both GGA-PBE and HSE functionals (using .1 $\AA ^{-1}$ screening length and 25\% HF-exchange mixing).}
\begin{ruledtabular}
\begin{tabular}{lcccc}
\textrm{functional} & \textrm{a (angs)} & \textrm{c/a} & \textrm{$E_g$ (eV)} & \textrm{$\Delta H_f$(eV)}\\
\colrule
GGA-PBE &8.108 & 1.408 & .32 & -1.79\\
HSE&8.160 & 1.390& 1.42 & -1.32\\
Experiment\cite{exp,kimball}&8.097 & 1.286& 1.49 & -1.53
\end{tabular}
\end{ruledtabular}
\end{table}

\section{\label{results}Results}
At room temperature and atmospheric pressure $Zn_3P_2$ forms a tetragonal phase with a 40-atom unit cell possessing symmetry P42/nmc.  It is derived from the cubic fluorite structure with zinc at the center of a distorted phosphorus tetrahedra and the phosphorus surrounded by eight zinc sites lying roughly at the corners of a cube, only six of which are occupied\cite{catalano} - see Figure \ref{unitcell}.  The calculated structural and basic thermodynamic parameters are summarized in Table \ref{Tab1}.  Results for the lattice constant and formation enthalpy are in good agreement with experiment with both GGA and HSE functionals.  However, the bandgap is severely underestimated with GGA, which at 0.32eV is only about 1/5 of the experimental result.  Using the HSE functional, we find the bandgap to within 5\% of experiment, highlighting the value of using HSE for this system.

\begin{figure}[t]
\caption{\label{bands}DFT Bandstructure - calculate with HSE functional.  Partial density of states shows makeup of VBM mainly of phosphorus p-character, while the CBM is mixed p and s-character from phosphorus and zinc states.}
\includegraphics[trim=0cm 3cm 1cm 3cm, clip=true,width=90mm]{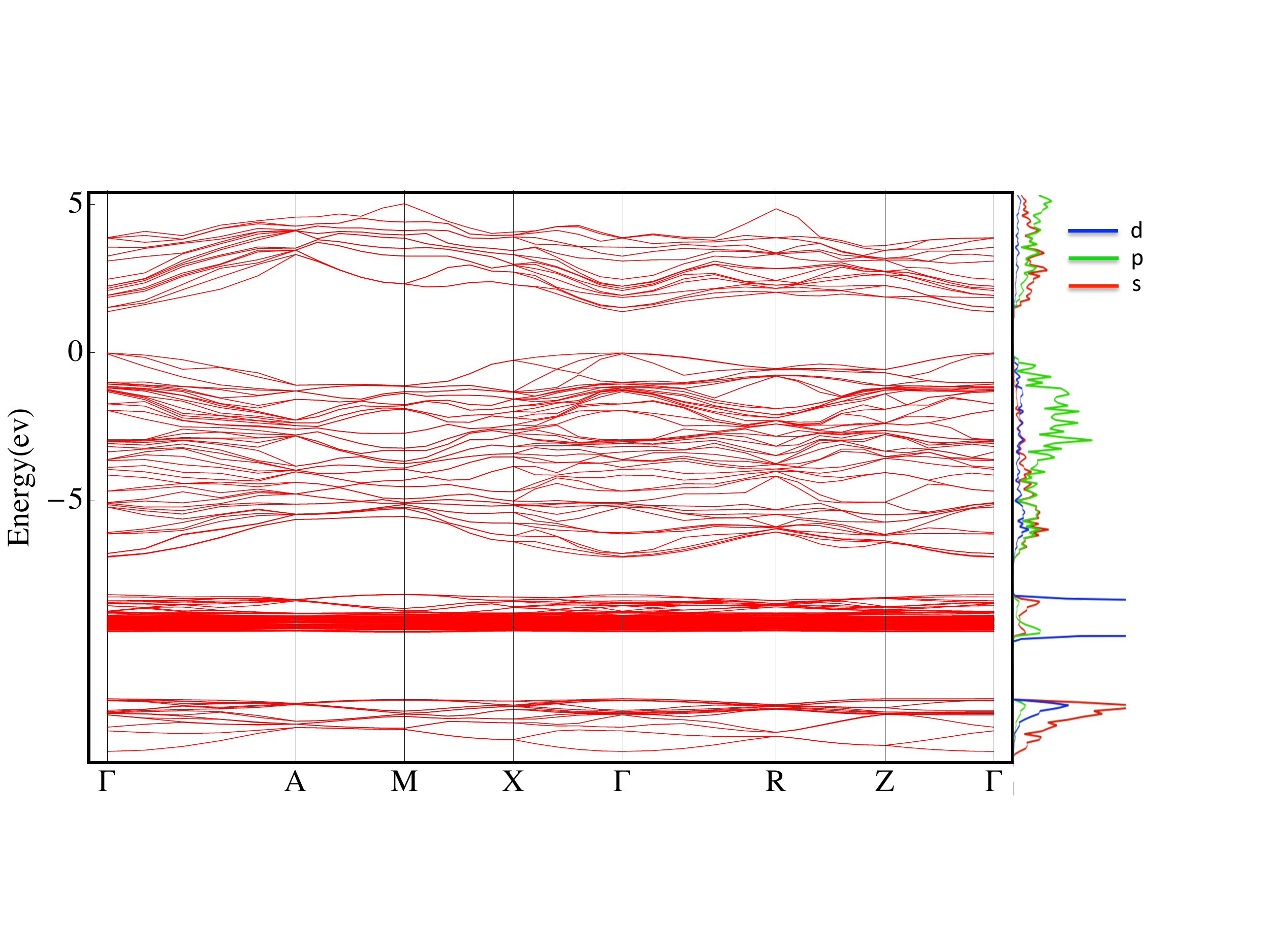}
\end{figure}

\begin{figure}[t]
\caption{\label{slab}Slab Geometry for band alignment determination.  Red regions show areas where the potential was integrated in the bulk and vacuum regions.}
\includegraphics[trim=0cm 1cm 0cm 2cm, clip=true,width=80mm]{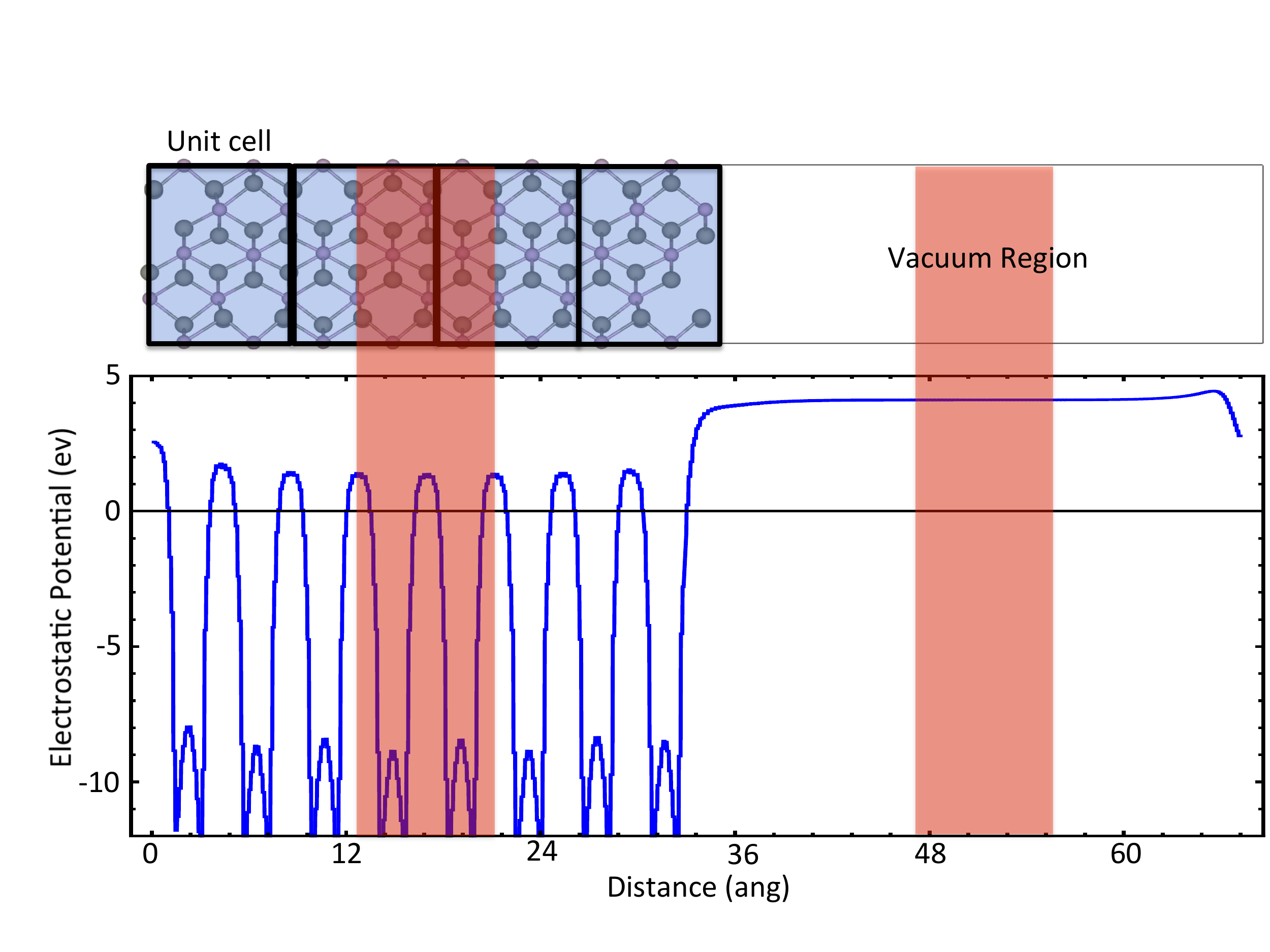}
\end{figure}

In order to accurately describe the defect levels we need to determine the effect of the HSE functional on the bandstructure (see Figure \ref{bands}) relative to GGA.  Calculating the band shifts from GGA to HSE requires more than simple bulk calculations, as the band energies given by DFT are only referenced to the average electrostatic potential of the simulation cell.  For a periodically repeated solid this is an ill-defined quantity which makes it impossible to directly compare band energies between cells with different constituents or for calculations performed with different functionals.  However, if we have a region of equivalent potential (ie. vacuum) in both GGA and HSE systems we can compare the bulk potentials in each system to this region, and subsequently compute the HSE and GGA band positions relative to each other.  Care must be taken that the solid and vacuum regions are large enough that the electron wavefunctions become negligible in the vacuum and the effects of the surface states are not felt in the bulk.  In this work, we use a 4 unit-cell slab along the non-polar [100] direction and an equal amount of vacuum (resulting in more than 35 $\AA$ of vacuum) - see Figure \ref{slab}.  Non-polar, non-reconstructed surfaces were used to avoid creating surface dipoles with the accompanying undesirable step in potential across the solid-vacuum interface.

\begin{figure}[t]
\caption{\label{bandalign}GGA to HSE Band Alignment.  Bandgaps for GGA (left) and HSE (right) are labeled at the gamma point.  Calculated offsets for the VBM and CBM are shown.}
\includegraphics[trim=1cm 3cm 4cm 3cm, clip=true,width=80mm]{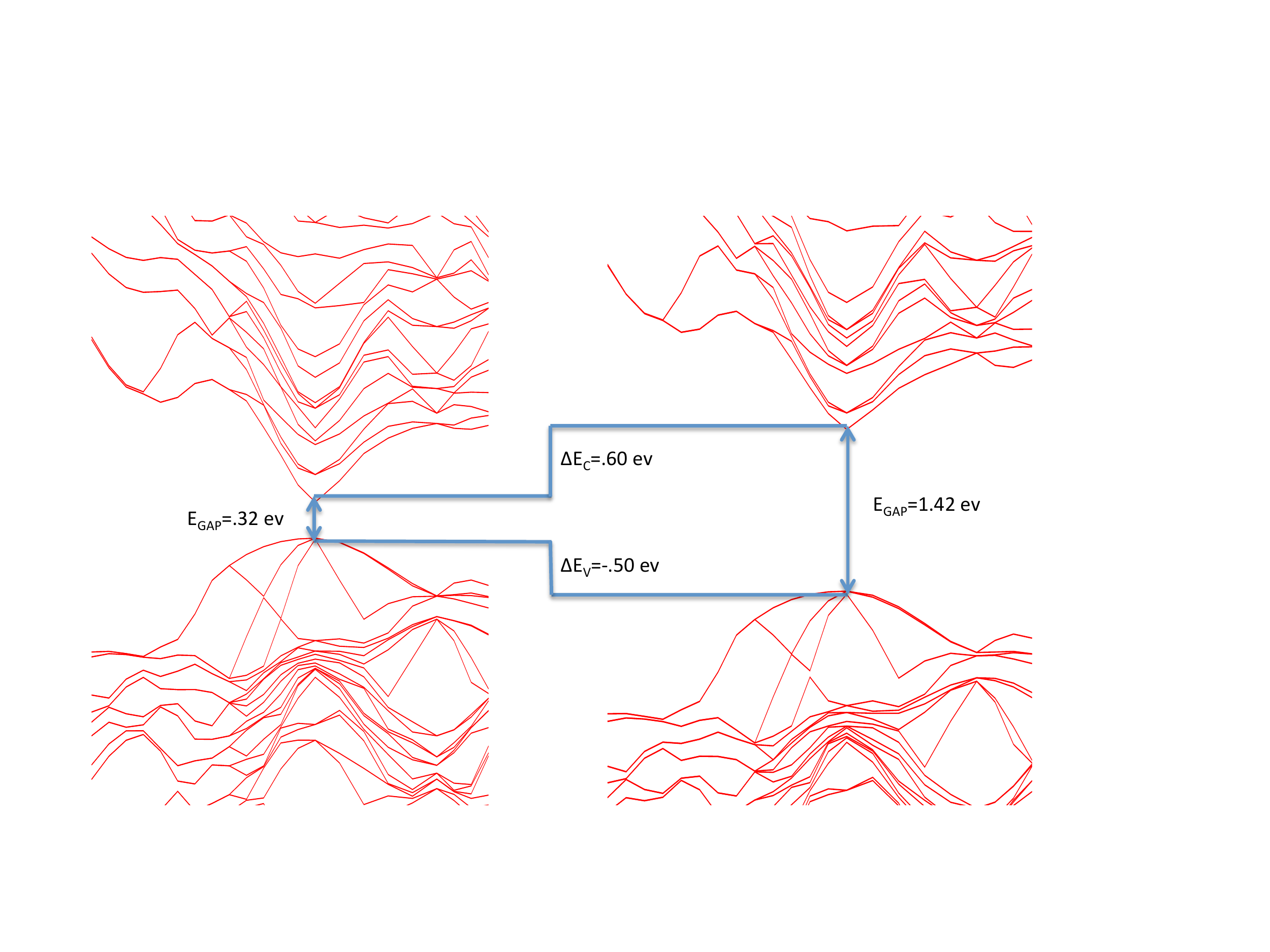}
\end{figure}

The alignment of GGA and HSE bandstructures is shown in Figure \ref{bandalign}.  The lowering of the VBM with HSE is due to the reduced self-interaction of the phosphorus p-orbitals which primarily form the highest-energy valence bands.  The upward shift of the CBM is due to the reduced hybridization of the Zn-s and P-s orbitals that make up the lower part of the conduction band with the valence band states.  With this alignment calculated we can compute the GGA-to-HSE band offsets needed for the 'perturbation extrapolation' method discussed in Section II B.

We are now in the position to apply all the corrections detailed in Section II.  We turn our attention to the predicted stability and doping effects of the various intrinsic defects.  Here it is useful to make a few notes about the following discussion; first, we may find that a defect level lies outside of the bandgap region predicting that this charged state is unstable as a defect localized state.  However, it may still be possible to bind electrons to the defect through electrostatic interation and form hydrogenic effective-mass states just within the band edge.  A discussion of these states is not in the scope of this work.  Furthermore, we ignore the effects of both formation volume and formation entropy in computing the defect formation energies.  Formation volume is directly related to the change in volume when a defect is created, but is typically only important for degenerate defect regime or for very high pressures.  Formation entropy for point defects is typically on the order of a few $k_B$ so are only important for very high temperatures - additionally, the entropy of point defects in the same system tend to be similar and so largely cancel when we compare the likelihood of one defect over another.

\subsection{\label{VZn}Zinc Vacancy}

\begin{figure}[t]
\caption{\label{DIST_VZn}$V_{Zn}$ Electronic Structure - Partial Density of States of the neutral defect for the four nearest neighboring phosphorus atoms.  Defect states highlighted in yellow, show mainly p-character.  Bond lengths are given relative to distance to defect site in perfect bulk cell.}
\includegraphics[trim=1cm 4cm 0cm 2cm, clip=true, width=80mm]{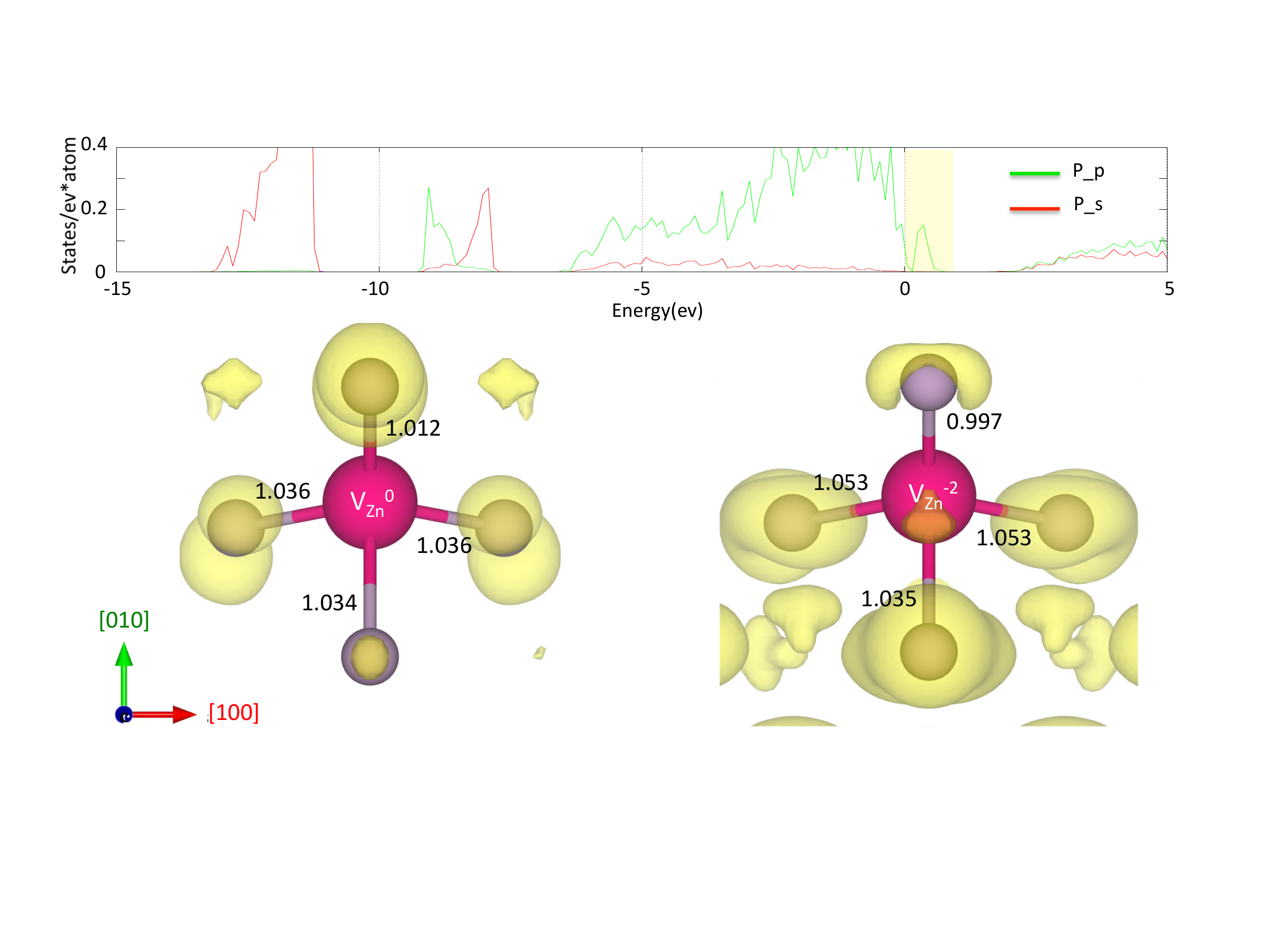}
\end{figure}

Zinc is surrounded by four nearly equidistant, tetragonally coordinated phosphorus atoms.  Removing a zinc atom from the lattice leaves four dangling bonds from the neighboring phosphorus atoms with mainly p-character - see Figure \ref{DIST_VZn}.  These bonds are occupied with only six electrons, forming three low energy and one empty higher-energy defect state.  This empty bonding state can capture electrons and form an acceptor defect with a -1 or -2 charge.  As the defect states have a strong valence band character (e.g. in terms of symmetry), we would expect shallow defect behavior.

It is instructive to study the structural relaxations around the $V_{Zn}$ site as this is intimately related to the occupancy and energetics of the defect state.  Removing a zinc atom to form a neutral defect causes the four neighbouring phosphorus atoms to relax away from the defect site as they seek to maximize their bond overlap with the remaining zinc atoms in the lattice.  However, this movement also shifts the dangling bonds in the vacancy region higher in energy, fighting this relaxation.  The phosphorus atom closest to the defect site (the P-atom to the 'north' of the vacancy site in Figure \ref{DIST_VZn}) has more of its density in the vacancy region and subsequently relaxes the least.  As the defect becomes occupied (as in the -1 or -2 charged states) the electrons are chiefly populating around the phosphorus ions which cause a further relaxation away from the defect due to increased coulombic repulsion.  As the relaxation is relatively slight in all defect charge states, we would expect that the defect itself would be delocalized.  This is born out as the charge density is poorly screened and well dispersed throughout the lattice.

\begin{figure}[t]
\caption{\label{FE_VZn}$V_{Zn}$ Formation Energy.  Acceptor-type defect, plotted for the 0,-1,-2 charged states as the fermi level varies from the top of the VBM (p-type) to the bottom of the CBM (n-type).  Favorable growth condition on the left.}
\includegraphics[width=80mm]{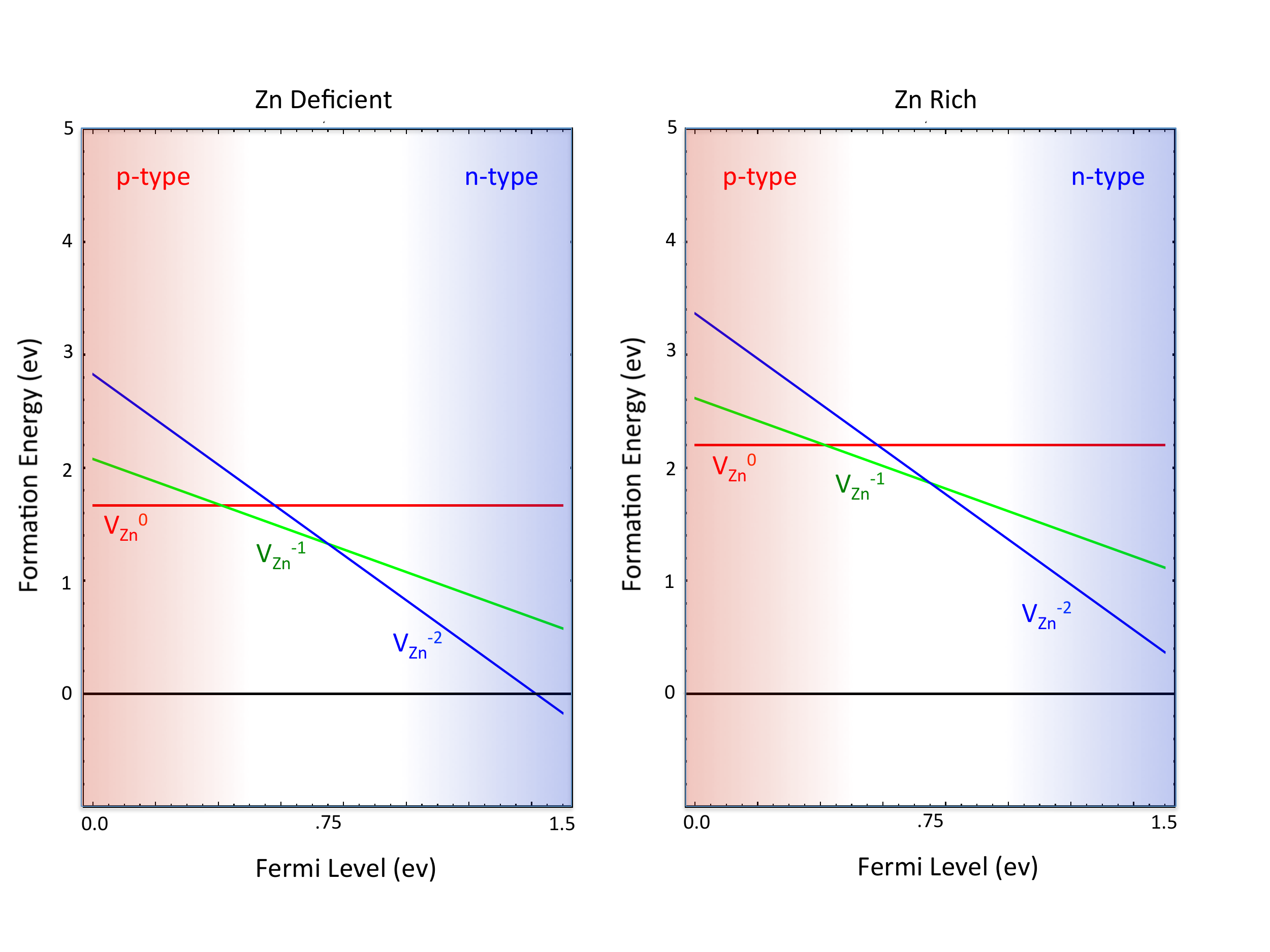}
\end{figure}

Getting the correct energetics and occupancy of the defect states is primarily important for their effect on the formation energy of the defects, which is our central goal.  Looking at the formation energy plots in Figure \ref{FE_VZn}, zinc vacancies exist as charged defects for all but very p-type regimes (fermi level close to VBM).  The small energy difference between the different charged states for a fermi level at the VBM can be expected from the small difference in lattice relaxations associated with the charged states of $V_{Zn}$.  The formation energies in the n-type regime are low enough in the Zn-deficient growth environment for this to be an important defect.\newline

\subsection{\label{VP}Phosphorus Vacancy}

\begin{figure}[t]
\caption{\label{DIST_VP}$V_P$ Electronic Structure - Partial Density of States of the +2 charged defect for the nearest neighboring zinc atoms.  Defect states highlighted in yellow, show mainly Zn-p and Zn-d character.  Bond lengths are given relative to distance to defect site in perfect bulk cell.}
\includegraphics[trim=1cm 2cm 0cm 2cm, clip=true, width=80mm]{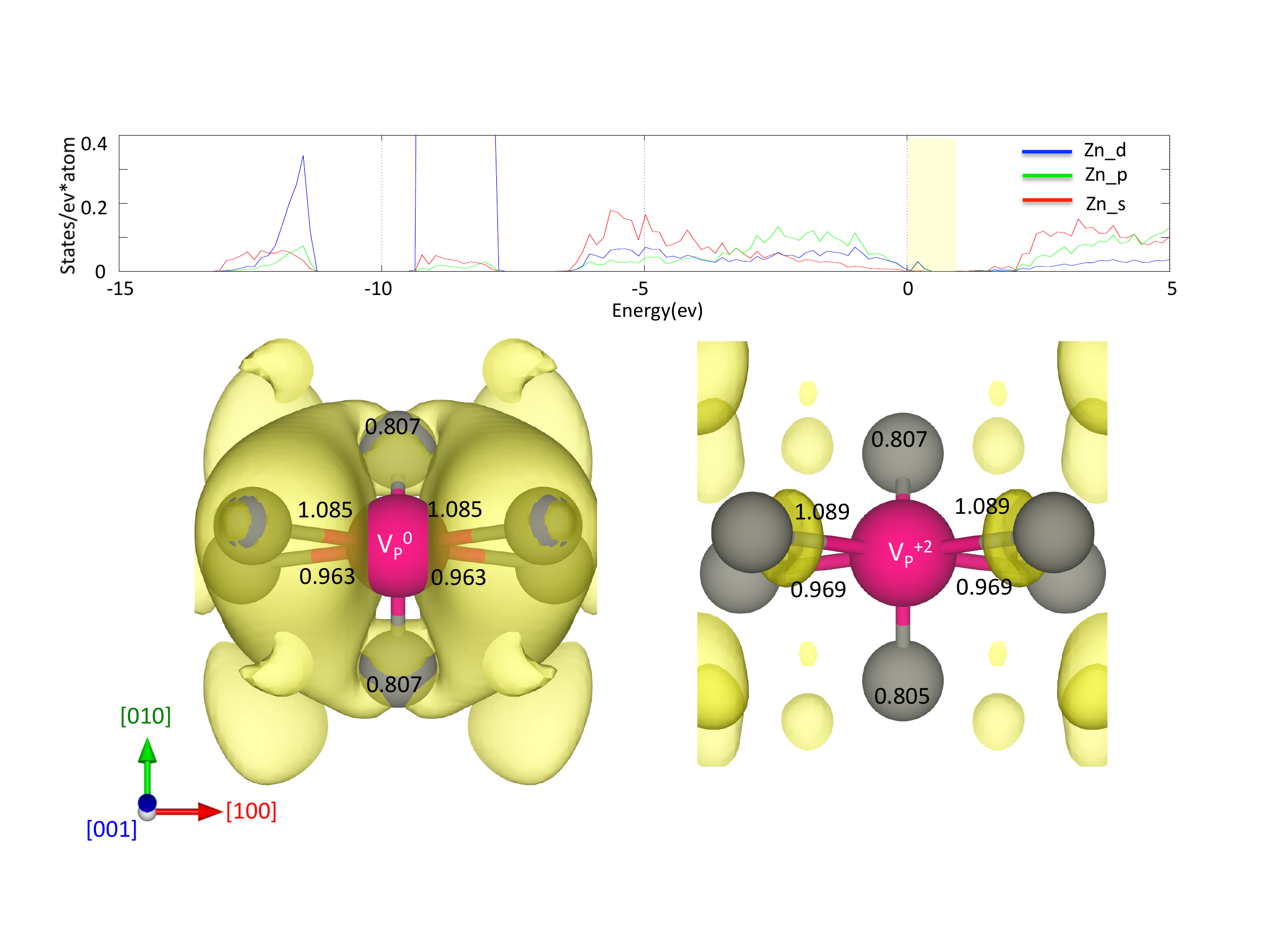}
\end{figure}

Phosphorus is surrounded by six nearest neighbor zinc sites, four being almost equidistant and two zinc atoms about 15\% further away.  Phosphorus normally exists in a -3 oxidation state and when we remove a phosphorus atom the electrons from the surrounding zinc atoms have only high energy bonding states to move into with mainly Zn-p and Zn-d character - see Figure \ref{DIST_VP}.  We would expect that these states would want to depopulate and form +1 and +2 charged defects as these electrons are no longer needed for bonding a phosphorus atom to the lattice.  The symmetry of the defect states is not similar to the conduction band symmetry so we would expect deeper defect behavior than for $V_{Zn}$.

Removing a phosphorus atom results in a defect state that is effectively screened by the neighboring zinc atoms and highly localized between the nearest zinc sites.  The high degree of localization results in a large lattice relaxation, especially for the two zinc atoms closest to the defect site.  Since the majority of the electron density is concentrated in the defect region, the four closest zinc atoms relax closer to each other in order to lower the energy of the occupied defect localized states.  The two furthest zinc atoms relax away from the defect region due to increased coulombic repulsion from the similarly charged zinc atoms closer to the defect.  As the defect becomes depopulated there is a slight outward relaxation since the benefit in lowering the energy of the defect states is reduced and the zinc atoms want to increase their bonding with the rest of the lattice.  The high degree of localization would also suggest deep defect behavior.  

\begin{figure}[t]
\caption{\label{FE_VP}$V_P$ Formation Energy.  Donor-type defect, plotted for the 0,+1,+2 charged states as the fermi level varies from the top of the VBM (p-type) to the bottom of the CBM (n-type).  Favorable growth condition on the left.}
\includegraphics[width=80mm]{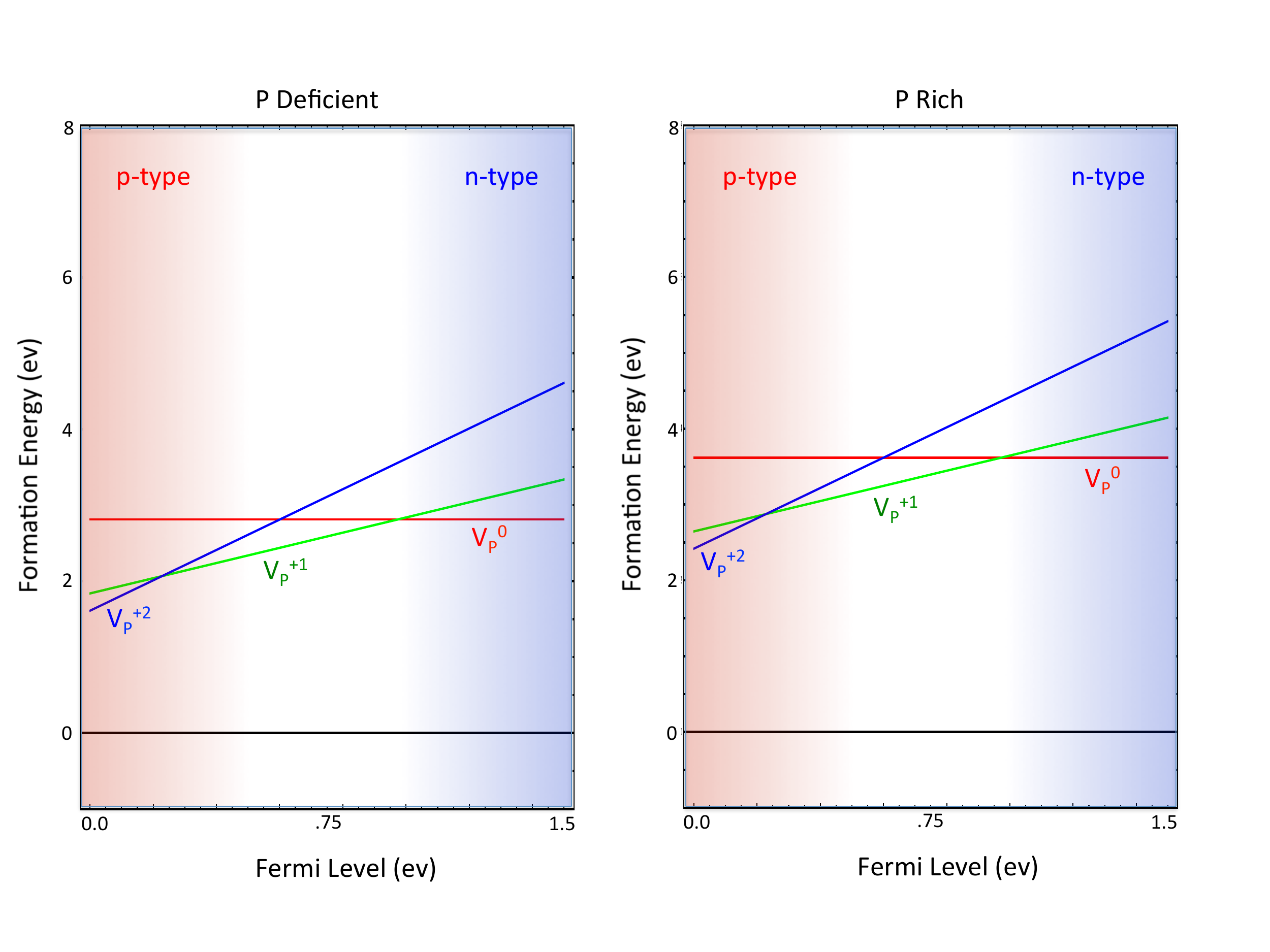}
\end{figure}

Looking at the formation energies in Figure \ref{FE_VP}, we see deep defect behavior where the defect only becomes charged for fermi levels in the neutral to p-type regime.  Even for favorable growth conditions these defects are too high in energy to play a significant role in the dopability of zinc phosphide.
\newline

\subsection{\label{IZn}Zinc Interstitial}

For both zinc and phosphorus interstitials we voxelized the $Zn_3P_2$ unit cell and tested the sites in order of furthest distance from neighboring atoms.  Zinc interstitials are predicted to form in the voids on the zinc plane of atoms tetrahedrally coordinated with four phosphorus atoms and in-line with zinc atoms in adjacent planes.  Incorporating an extra zinc atom into the lattice forms a localized defect state of mainly ionic Zn-s character since the neighboring phosphorus atoms have closed-shell configurations.  As for zinc atoms in the perfect bulk, the interstitial zinc has a tendency to depopulate the s-shell states.  We would expect donor defect behavior with a +1 or +2 charged state and this is what our calculations show.  The symmetry of the defect states are similar to the conduction band character, consequently we would expect shallow behavior.

The neighboring zinc atoms are affected the most by the incorporation of the $Zn_i$ defect as they relax away from the similarly charged interstitial ion.  The phosphorus atoms relax closer to the interstitial site in general as there has been a partial charge transfer from the $Zn_i$ defect.  There is only minor differences in relaxation between the various charged states and the defect is poorly screened with a delocalized wavefunction centered on the majority of the zinc atoms in the lattice.  All of which would suggest shallow defect formation.

\begin{figure}[t]
\caption{\label{FE_IZn}$Zn_i$ Formation Energy.  Donor-type defect, plotted for the 0,+1,+2 charged states as the fermi level varies from the top of the VBM (p-type) to the bottom of the CBM (n-type).  Favorable growth condition on the left.}
\includegraphics[width=80mm]{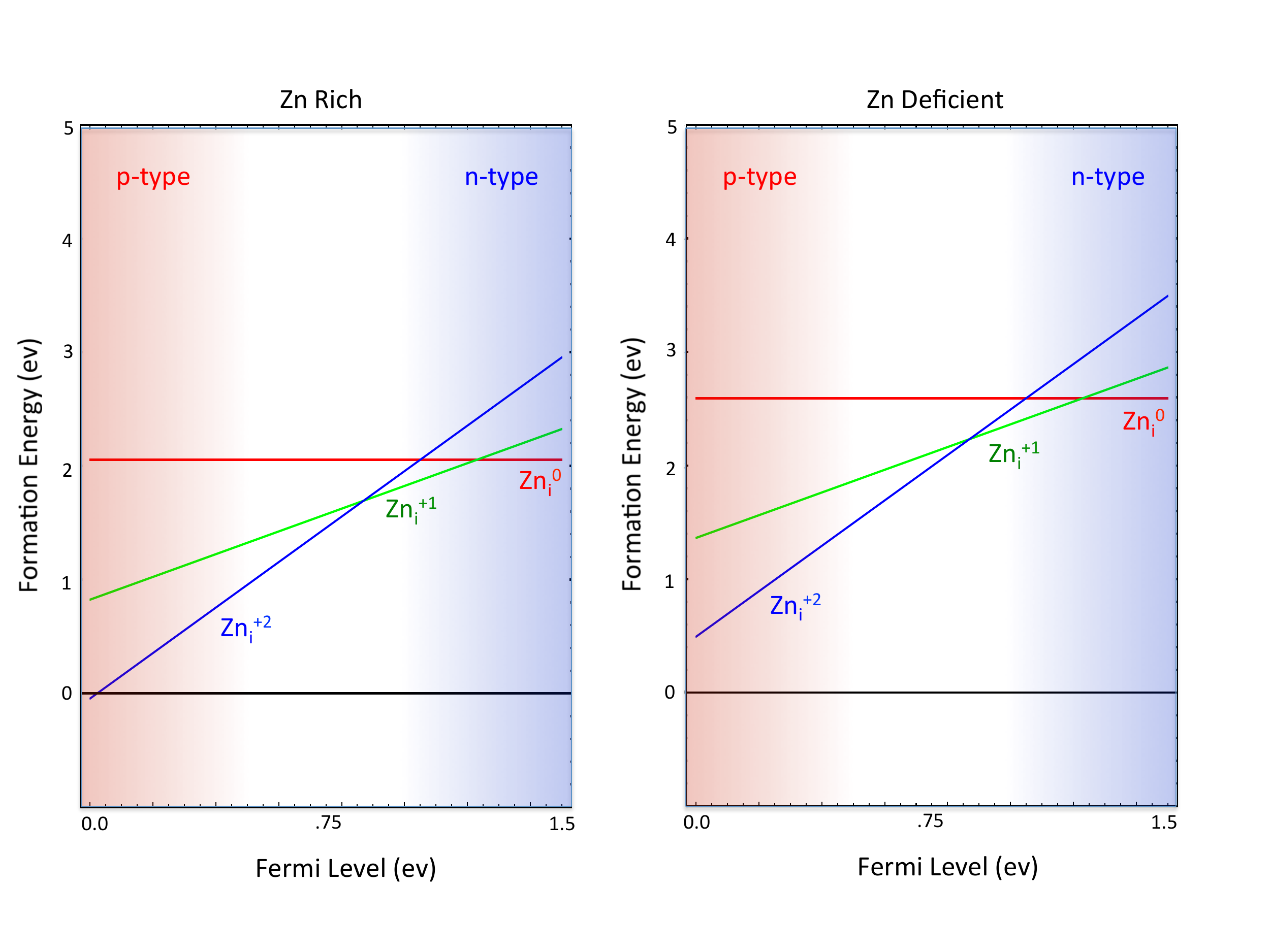}
\end{figure}

In Figure \ref{FE_IZn}, the formation energy plots show shallow behavior where the defect becomes charged for all but fermi levels high in the n-type regime.  As the fermi level drops to the VBM the formation energy of the defects becomes very small for the favorable growth conditions (Zn-rich regime).  These defects should be important to consider.
\newline

\begin{figure}[t]
\caption{\label{DIST_IZn}$Zn_i$ Electronic Structure - Partial Density of States of the -2 charged defect for the interstitial Zn site.  Defect states highlighted in yellow, show mainly Zn-s character.  Bond lengths are given relative to distance to defect site in perfect bulk cell.}
\includegraphics[trim=1cm 2cm 0cm 2cm, clip=true, width=80mm]{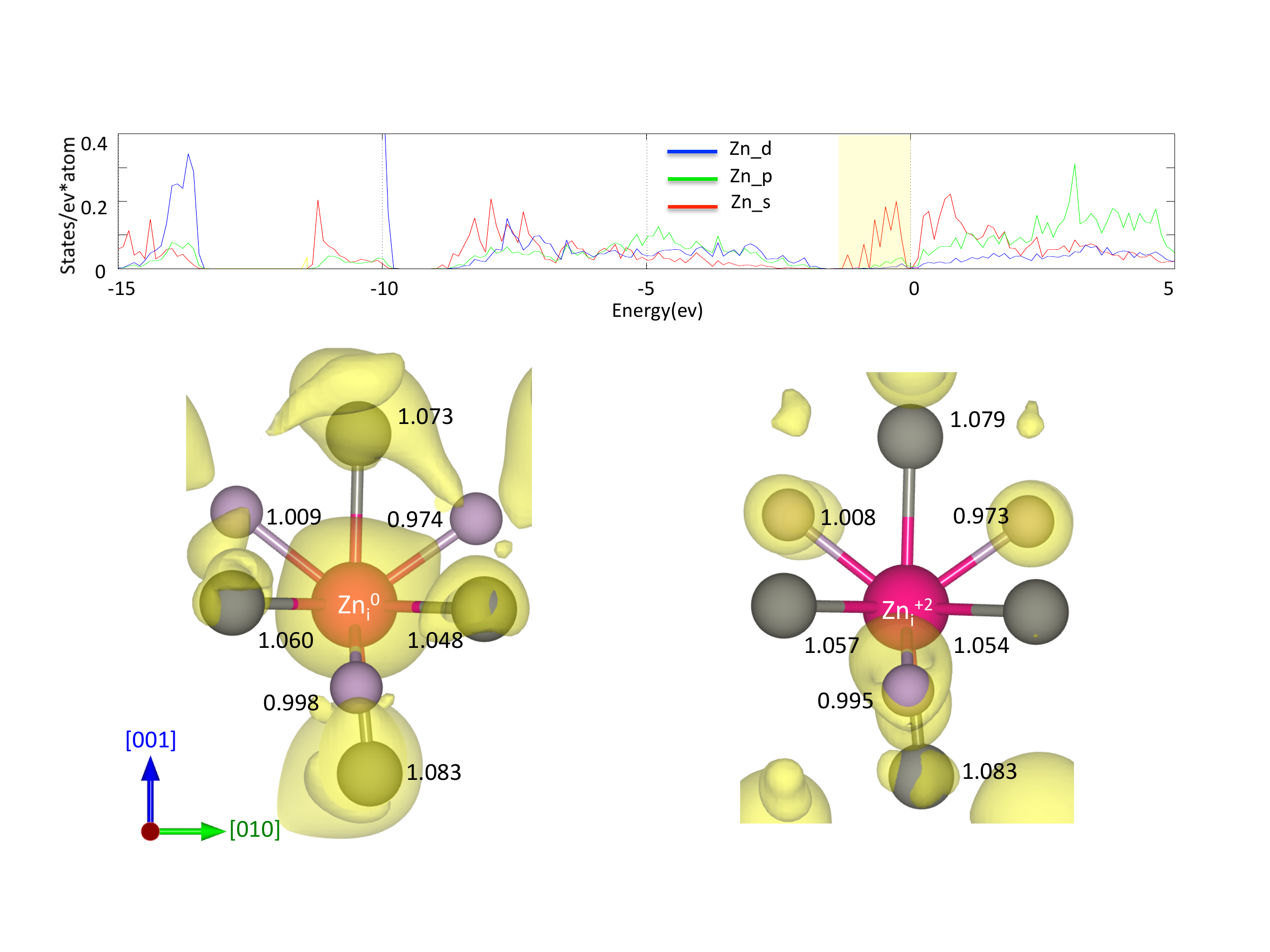}
\end{figure}

\subsection{\label{IP}Phosphorus Interstitial}

\begin{figure}[t]
\caption{\label{DIST_IP}$P_i$ Electronic Structure - Partial Density of States of the neutral defect for the interstitial phosphorus site.  Defect states highlighted in yellow, show mainly P-p character.  Bond lengths are given relative to distance to defect site in perfect bulk cell.}
\includegraphics[trim=1cm 2cm 0cm 2cm, clip=true, width=80mm]{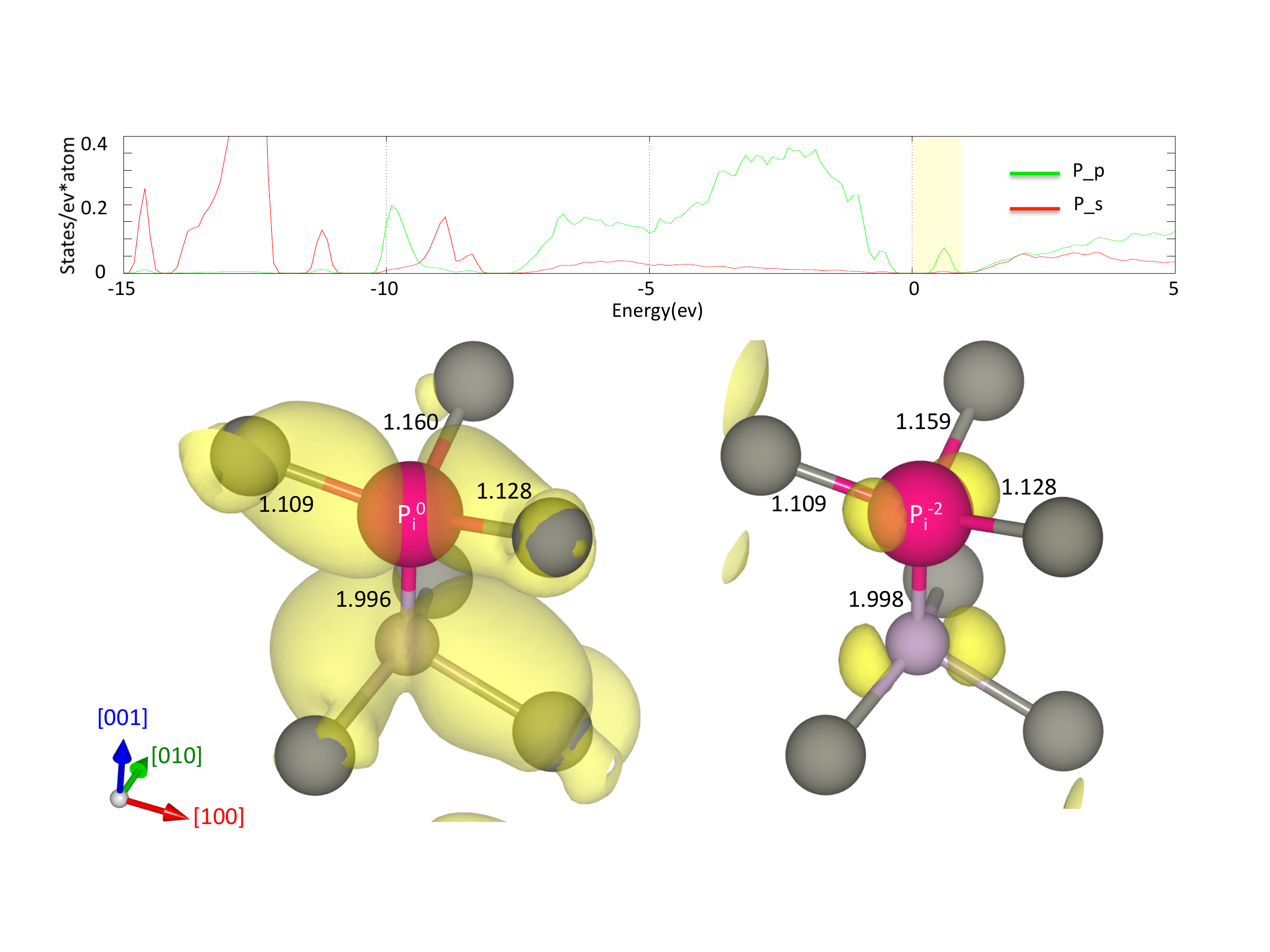}
\end{figure}

The most energetically favorable position for the phosphorus interstitials are in the voids in the zinc plane of atoms nearly equidistant from three zinc atoms and directly above a phosphorus atom displaced from the plane below.  The additional electrons form states of mainly covalent p-character between the $P_i$ site and neighboring zinc and phosphorus atoms - see Figure \ref{DIST_IP}.  Due to the natural oxidation state of phosphorus (-3) we expect that the defect should be an acceptor type, which is confirmed by our calculations.  Since the symmetry of the defect states is similar to the VBM, we would predict shallow defect behavior.

The neutral defect state is highly localized, so $P_i$ causes relatively large ionic relaxation away from the defect site, though for the two zinc atoms most involved with the covalent bonding this relaxation is reduced because this movement also increases the energy of the populated defect states.  The charged state is poorly screened and very delocalized across the lattice.  Consequently, there is very little relaxation as the defect captures electrons from the conduction band.  Apart from energetics, the severe delocalization of the charged state and good symmetry match between the defect and valence states causes almost band-like behavior where we should readily capture electrons from the conduction band.

The neutral defect has relatively low formation energy for P-rich growth conditions.  Due to the small differences in relaxation between the charged defect states, there is only a small difference in formation energy between them at the VBM.  Thus, for fermi levels even modestly into the n-type regime the formation energy of the charged defects becomes very small making these critical defects to consider.\newline

\begin{figure}[t]
\caption{\label{FE_IP}$P_i$ Formation Energy.  Acceptor-type defect, plotted for the 0,-1,-2 charged states as the fermi level varies from the top of the VBM (p-type) to the bottom of the CBM (n-type).  Favorable growth condition on the left.}
\includegraphics[width=80mm]{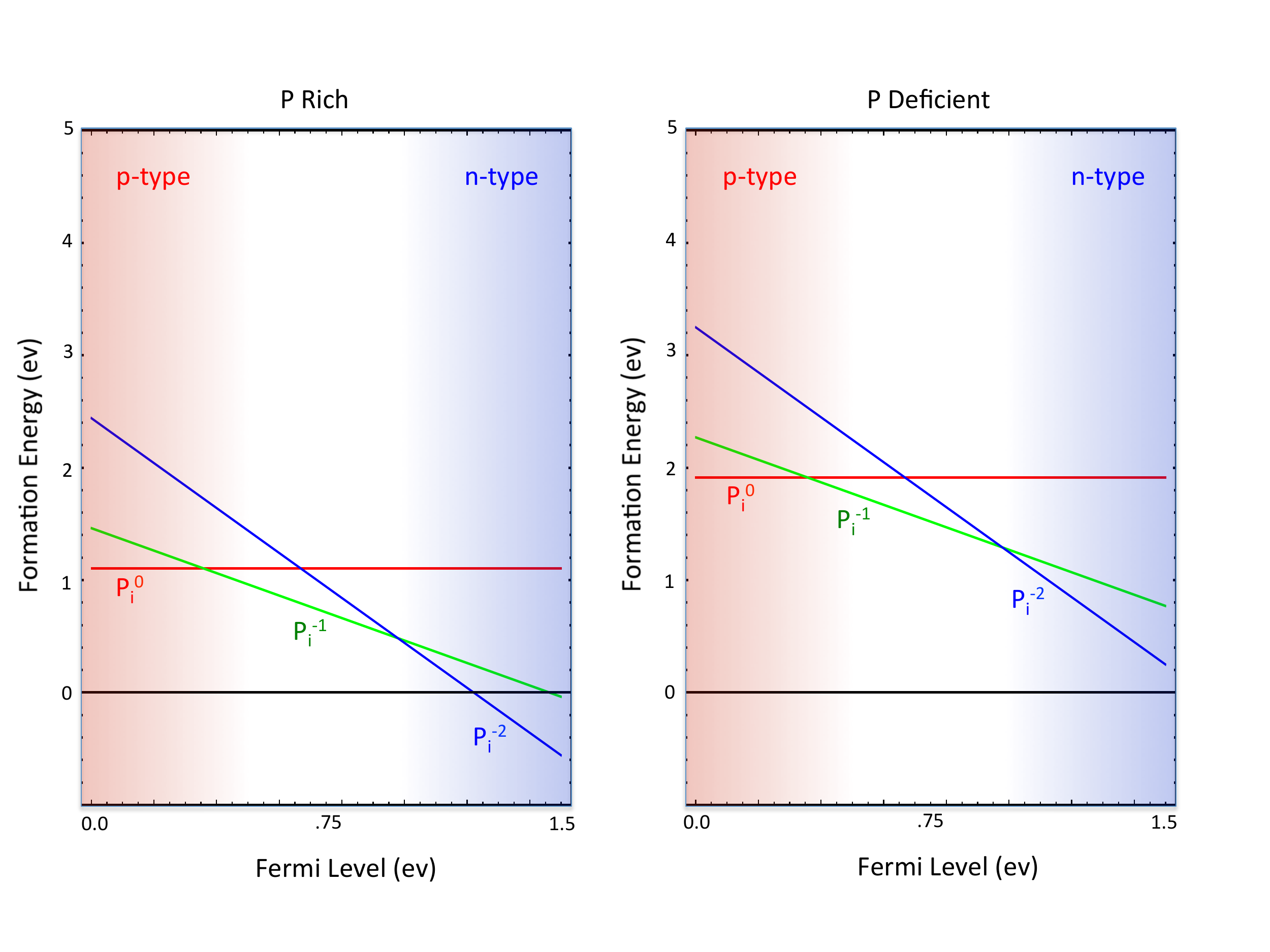}
\end{figure}

\subsection{\label{conclusion}CONCLUSION}
From our calculations, the likely candidate for the lack of n-type dopability is the $P_i$ defects.  Both of the acceptor defects ($V_{Zn}$ and $P_i$) have low formation energies as we move the system into the n-type regime.  While there are more suitable $V_{Zn}$ sites, the $P_i$ defect are significantly less costly and should be the vastly more prevalent defect.  The $Zn_i$ defects would tend to aid n-type doping as they are electron donors, though they have too high a formation energy for fermi levels even moderately n-typed to significantly compensate for the $P_i$ defects.

Thus, as we dope our system with electrons we create a large amount of acceptor defects which act as 'electron-sinks' and capture mobile electrons from the conduction band and neutralize the doping.  In fact, the $P_i$ defect requires zero formation energy for fermi-levels midway towards the conduction band.  This would 'pin' the fermi-level - that is, as we try to approach this level of doping a massive amount of $P_i$ defects would form in the crystal.  Since the creation of $P_i$ defects fight n-type doping, reaching this level would not be expected.  There is some hope in that we would anticipate the $P_i$ defects to
repel each other (especially the highly charged state), thus limiting their concentration, and
some n-type doping would then survive.  Indeed, initial results show a significant penalty in having two fully charged $P_i$ defects within the same 2x2x2 supercell.  However, this still corresponds to a large defect concentration and there is low likelihood that the desireable properties of $Zn_3P_2$ such as high minority carrier diffusion lengths would survive in the regime where $P_i$ defects saturate.

Solving for the fermi level self-consistently using the above formation energy functions of the various defects, we predict an intrinsic fermi-level of 0.55 eV, which is mildly p-type.  Pushing the fermi level beyond roughly 1.05 eV becomes impossible as the number of defects exceeds the numer of sites in the cell at this point.

Since the $P_i$ defects are the problematic defects, any means to suppress them should help the n-type doping issue.  For extreme Zn-rich growth conditions, where the acceptor $P_i$ defects are suppressed and the donor $Zn_i$ are enhanced there may be some hope of weakly n-type materials being formed.   A suppression of interstitial defects in general, such as straining the crystal as it is grown may prove fruitful.  Cluster doping\citep{Ligen} donor atoms with ones that suppress $P_i$ formation would be another avenue to explore.

\begin{acknowledgments}
Thanks to Qijun Hong for help in processing VASP PAW wavefunction files.  Fruitful discussions with Prof. Harry Atwater, Jeff Bosco, and Greg Kimball are gratefully acknowledged.  This work made use of the TeraGrid/Xsede computing resources at the University of Texas at Austin.
This material is based upon work supported by the Department of Energy National
Nuclear Security Administration under Award No. DE-FC52-08NA28613
and by the National Science Foundation under Grant No. DMR-0907669.
\end{acknowledgments}

\bibliography{ZnP_paper_v4}
\end{document}